# FROM MY PERSPECTIVE
# The Sad State of Entrepreneurship in America: What Educators Can Do About It

Fred Phillips, Stony Brook – State University of New York (fred.phillips@stonybrook.edu)[1]

While responsible people work to implement the UN Sustainable Development Goals, and to stave off humanity's ecological suicide, young nerds in Silicon Valley write unimportant apps to do (as one genius Internet meme put it) "things their mothers used to do for them." The Elon Musks of the world – those who prepare for the long game while financing it with innovative products for today's market – are so rare as to be anomalies.

Today's entrepreneurial scene suffers from a sick venture capital industry, a number of imponderable illogics, and, maybe, misplaced adulation from students and the public. The ailments also include:

- "Frat-boy" start-up cultures that waste money and denigrate women and minorities (Wadwha, 2014).
- Venture capital investors (VCs) with no management experience, attempting to micro-manage the companies they invest in.
- VCs who prefer to invest in – and then manipulate and bully – young entrepreneurs, when in fact more successful new firms are started by older entrepreneurs with corporate experience (Kauffman Foundation, 2017).
- Engineer-entrepreneurs with no knowledge of humanities or liberal arts – i.e., with no idea of what products society needs, or how their development projects will fit into societies and markets.
- Insufficient attention to data security. E.g., Snapchat (Ribeiro 2014) and MongoHQ (TrendMicro 2013).
- In many states and countries, government startup funding going to politically favored but unqualified entrepreneurs.
- Outright frauds like Theranos[2] and Pathway Genomics (Duhaime-Ross 2015).

At present, 30% of all entrepreneurs in the U.S. are foreign-born and nearly 80% of all the intellectual property in the U.S. in the high-tech sector – telecomm, semiconductors, and life sciences – has its origins in foreign-born entrepreneurs (Angelos Angelou, quoted in Newlands, 2017). America is, at present, the world's font of innovation. As Bernie Sanders and Michio Kaku have both noted on YouTube, this flow of innovation depends on immigrant brains and initiative. Headlines like "Is Brand America Tanking?" in the *Forbes* business magazine anticipate an America that is less attractive to entrepreneurial migrants, and thus a less vibrant innovator.

New ventures seem focused on growth (in order to give VCs a quick and rich liquidity event), not on delivering superior customer value. Luckily, worthless online advice to startups like "Put your resources into customer acquisition over product" (Prajapati 2017) is balanced by more sensible pundits. Anand Sanwal of CBInsights, for example, calls out the "bullshit culture" of catering to investors rather than to customers.[3]

---

[1] The author thanks Mr. Be Mix Le for valuable research assistance.
[2] https://www.theverge.com/tag/theranos
[3] http://us1.campaign-archive.com/?u=0c60818e26ecdbe423a10ad2f&id=ac6cb97638&e=3c60089143

*Venture capital*

Lazy VCs know they need to tap the innovation potential of smart people who do not live in San Jose, Seattle, or Austin, but they still insist that companies they invest in must move to where the VCs are.[4] Opportunities are lost. Civic-minded entrepreneurs, who want their companies to benefit their hometowns, are out of luck if their hometowns are St. Louis or Flagstaff. Despite a flowering of innovation in Asia, for example, few US VCs are willing to invest in overseas startups.[5]

The unregulated US VC industry[6] attracts, well, people who are attracted to unregulated financial markets, with all that that implies. Suffice it to note that VCs spout insider jargon ("space," "unicorn," "pivot") to obscure the fact that they really don't know anything. They invest in what other VCs invest in – one entrepreneur called VCs "lemmings" – hesitating to risk money outside the box.

Their culture of morphing failure into virtue ("I'd invest in him again because now he's experienced; he knows how the game is played"), possibly justified in the past, has metastasized. Now no one is responsible for anything. Fisher (2017) calls Silicon Valley the "land of no consequences":

> "Venture capital comprises such a small portion of the portfolio of large [pension funds] that there is little impetus to care about where VCs distribute cash.
> "In the later stages, a startup will go public or be acquired and, even if it is horrible, the VCs… will cash out, leaving others—your mom and dad's pension fund—to deal with the fallout.
> "Where is the [institutional investor] who is coming forward and actually withdrawing money from a sketchy VC? Where is the VC who admits they propped up an unsustainable company—along with underwriters—in order to make money and flip the risk to the public market?"

VCs exert unreasonable pressure on their startups, implying that a less than billion dollar valuation, or not disrupting a trillion dollar market, is essentially failure. "Stratospheric expectations are killing fledgling startups," says Wickre (2017).

More than one successful entrepreneur has told me, "I will never start a VC-backed company again." I must add that I am acquainted with a few highly intelligent VCs. But I am convinced they are members of a very small minority.

*Strange contradictions*

One of the contradictions of today's entrepreneurship scene, then, is that failure is simultaneously glamorized and forced. If an investor said to you, "You're not going to be a unicorn, we are withdrawing our money and shutting your company, sorry, come back again with your next idea," would you not recognize the con?

VCs perform financially no better than stock index funds, but with higher risk (Mulcahy et al. 2012). Chamath Palihapitiya, founder of VC firm Social Capital, goes so far as to call VCs "worthless" (CBInsights 2017). He claims VCs are motivated "to get credit, [to get] a *TechCrunch* article, to get a press release." People are rewarded for "making good [Powerpoint] decks… not creating value." VC Eric Paley (2017) is equally blunt:

---

[4] VC Dave McClure, interviewed by Caroline Fairchild in 2016 at https://www.linkedin.com/pulse/500-startups-founder-venture-capitalists-lazy-dont-caroline-fairchild/

[5] http://startuplawyer.com/venture-capital/flipping-your-international-startup-for-us-venture-capital

[6] See Mifsud et al (2010); Jickling and Murphy (2009). Singapore, in contrast, regulates its VC industry (Lim et al 2016).

> "Venture capital should come with a warning label. In our experience, VC kills more startups than slow customer adoption, technical debt and co-founder infighting — combined. VC should be a catalyst for growing companies, but, more commonly, it's a toxic substance that destroys them. VC often compels companies to prematurely scale, which is typically a death sentence for startups."

Nonetheless − in another strange contradiction − the city of Albuquerque dangled a $15 million lure to bring a California VC firm to New Mexico.[7]

A third contradiction: With its $93 billion Vision Fund, SoftBank has become a leading power in VC investment. Although IPOs have tended to benefit VCs more than ordinary investors (see above), "the big fear voiced by some analysts is that SoftBank's large check book will help firms stay private longer, which would be bad for the IPO market" (Popper and Lopatto, 2015). This fear is groundless, for the sad reason that other VCs have already made it happen. (See the McClure interview noted in footnote 3.)

Moreover, the publically touted valuations of unicorns and near-unicorns are at odds with these companies' "true" valuations after preferred stockholder perquisites are accounted for. Gornall and Strebulaev (2017) write, "The average unicorn is worth half the headline price put out after each new valuation." The level of small print in which these perks are hidden smells of potential financial fraud.[8]

A unicorn's valuation − if not the value − may still be high. Nonetheless, there ensues a fourth contradiction: The unicorns are not unicorns because of their intrinsically high value.[9] Low interest rates, according to the editors of *Verge*, have given us an "ever-expanding bubble in startup valuations, fueled by an ever-expanding pool of increasingly less qualified investors…. [Due to] pitiful or negative interest rates… people with a lot of capital will pay almost any price for the chance to earn a meaningful return" (Popper and Lopatto, 2015). Indeed, "after five years of ever larger funding rounds, the market threw cold water on the party. From Dropbox to Square to Snapchat, a lot of 'unicorns' failed to live up to their sky-high valuations." Softbank's huge investments may create another bubble, as the companies the Vision Fund invests in experience significant "valuation bumps."[10] Nikkei Asia fears that Asia's exchanges will lower their quality standards in order to attract more tech IPOs.[11]

*Healthcare: Botching the opportunity*

My students in healthcare management refused to sit in the same classroom with students from high-tech. "We are in a caring profession," they said, "The tech guys are looking only for profit. Talking with them is thankless." For their part, VCs early in this century shied away from healthcare because of the delays attached to FDA approvals. The culture gap has not narrowed in the ensuing twelve years. Again from the dialog of Popper and Lopatto (2015):

> What many firms seem not to understand is that though there are massive revenue opportunities in health care, "disrupting" patients' lives can lead to death. "Ask forgiveness, not permission" works fine in software. The medical field doesn't move as fast as the software industry because moving fast and breaking things is fine for things but not for people. The job of slow-moving bodies like the FDA is to keep companies from harming patients in their quest to get rich.

---

[7] https://www.abqjournal.com/1075943/nm-lures-silicon-valley-venture-firm-with-15-million-commitment.html

[8] Gornall and Strebulaev's research was picked up by the *New York Times* and other news outlets.

[9] http://us1.campaign-archive1.com/?u=0c60818e26ecdbe423a10ad2f&id=83da300a37&e=3c60089143

[10] Ibid.

[11] https://asia.nikkei.com/magazine/20171019

The thing is, I'm not sure Silicon Valley sees the difference…. [Healthcare and life sciences are] undeniably a slower, riskier investment than the next photo-sharing app…. Disruption is more dangerous when it comes to medicine.

*What should educators do?*

1. *Before students are admitted.* Campaign to improve pre-college education in the USA, including science curriculum. Join the AAAS and other concerned organizations to push for public and governmental respect for the scientific method and its conclusions. Ally with high school educators, so that students will be exposed to these concepts for eight years rather than just four. Admit only students whose application essays show concern for people, planet, and profit.

2. *While students are enrolled.* Emphasize responsible innovation in the university curriculum. Teach the ethical, legal, and technical aspects of data security. Include more liberal arts courses in business and engineering curricula. Do not allow students to participate in venture competitions unless their business plans show concern for people, planet, and profit.

The high-pressure life of founders has led to weight gain, failing personal relationships, depression, and even suicide (Velayanikal 2017). Weekly beer blowouts at the incubator are not enough to counter this. Young entrepreneurs should be taught practical skills for dealing with stress and overwork.

3. *Alumni and executive education.* Celebrate not just alumni who have made bundles of money (and are thus prime prospective donors to the U) but also those who have shown responsibility in their careers. Offer executive education to VCs. Make sure they are informed of the consequences of their actions.

Our enthusiasm for entrepreneurship began in the late 1970s when a researcher noticed that most job growth came from newer and smaller companies, not from the giant conglomerates of that era. As job growth is a prime objective of almost every city, state, and nation, it seemed obvious that entrepreneurship should be encouraged. Then the glamor took over, the high drama of VC investment, and the sudden vast riches. We began to idolize tech entrepreneurship for its own sake.

Are new and small companies still the font of job creation? The Kauffman Foundation (Casselman 2017) and the *Verge* editors (Popper and Lopatto, 2015) think so. From the latter (though they might be including mature venture-backed firms like Google): "When you look at the impact of public companies created by venture capital, they account for a disproportionate amount of the jobs created…"

Yet it is our job as academics and policy makers to check that this remains true. VCs invest to make money, not to create jobs. On the contrary, they enjoin their entrepreneurs to minimize headcount. As robots and artificial intelligence become better substitutes for human heads, it is all the more important to continually re-check the role of startups in job creation.

*Closing remarks*

There are some encouraging signs and some discouraging ones.

The US Federal Reserve appears to be ready to raise interest rates. This should help stanch the flow of "dumb money" into tech, life sciences, and healthcare.

The Sasin Graduate Institute of Business Administration of Chulalongkorn University now sponsors a venture competition purposed to "bring sustainability concerns to mainstream commercial venture[s]." An acceptable entry must be a "business plan that is proactive in sustainable development and innovation."[12] Winners are eligible for HM the King of Thailand's Award, and HRH Princess Maha Chakri Sirindhorn's Sustainability Award. Another new initiative,[13] supported by major funders including the Long Now Foundation, aims for "a more ethical and inclusive movement to counter existing start-up and venture capital culture." Two hundred funders and founders met in November, 2017 in Portland to kick off the initiative.

Partly as a result of the forces discussed above, the proportion of USA companies less than a year old has dropped from 15% in 1980 to 8% in 2017, according to the US Census (Casselman 2017). Casselman also attributes the decline to a deteriorating relationship between big companies and startups – a problem that, with Korean companies, has been recognized for years (Li et al 2016; Economist Intelligence Unit 2017; Premack 2017), but to date has remained under the radar as regards US companies.[14] Fortunately, Deloitte is taking steps to attempt improvement in these relationships (Cheong, 2017).

I have mentioned the great efforts spent to create apps that make life infinitesimally more convenient. At a much higher level of wasteful insignificance, we have the rich who are afraid to die, and who are funding startups researching human longevity and even immortality – everything from cryogenics to uploading of human personalities into AIs. This may be useful if we ever attempt interstellar travel, but isn't it too soon to worry about it?

There is an easy way to extend average human life expectancy: Make sure underserved populations get adequate diet, sanitation, and healthcare. Again, Popper and Lopatto (2015) state the matter well: "None of this [reducing privilege gaps] is quite as sexy as living forever.… How do you monetize serving the poor? These companies grow rapidly at the expense of the overall health of the nation."

Fighting these dangerous trends is a formidable task for educators, but we must try.

*References*

Ben Casselman, A Start-Up Slump Is a Drag on the Economy. Big Business May Be to Blame. *New York Times*, Sept. 20, 2017. https://www.nytimes.com/2017/09/20/business/economy/startup-business.html

CBInsights, Why Startups Need To Go After 'Hard Things'. January 12, 2017. https://www.cbinsights.com/research/social-capital-vcs-healthcare-long-term-investing

Hui Min Cheong, Deloitte launches innovation tour to foster engagement between startups and corporates. *Tech in Asia*, Oct 13, 2017. https://www.techinasia.com/deloitte-launches-innovation-tour-foster-engagement-startups-corporates

Arielle Duhaime-Ross, Startup claims its test finds cancer early, but where's the evidence? *The Verge*, Sep 10, 2015. https://www.theverge.com/2015/9/10/9299603/pathway-genomics-cancer-blood-test-cancerintercept-evidence

The Economist Intelligence Unit, *South Korea: Making up for lost time. 2017, pp7-8.*

Yoav Fisher, Silicon Valley has become the land of no consequences. *Tech in Asia*, July 14, 2017. https://www.techinasia.com/talk/silicon-valley-no-consequences.

---

[12] http://bbc.sasin.edu/rules.php
[13] www.zebrasunite.com
[14] A giant Illinois company abused my own small consulting firm.